\documentclass[runningheads]{llncs}
\usepackage{graphicx}
\usepackage{epstopdf}
\usepackage{cite}
\usepackage{amsmath,amssymb,amsfonts}
\usepackage{xcolor}
\usepackage{algorithm}
\usepackage{algorithmicx}
\usepackage{multicol}
\usepackage{subfigure}
\usepackage{bbding}
\usepackage[noend]{algpseudocode}
\usepackage{booktabs}
\usepackage[colorlinks,linkcolor=blue,anchorcolor=blue,citecolor=blue]{hyperref}
\begin{document}
\title{Multi-owner Secure Encrypted Search Using Searching Adversarial Networks}
\author{Kai Chen\inst{1}\and
Zhongrui Lin\inst{2} \and
Jian Wan\inst{2}\and
Lei Xu\inst{1}\and
Chungen Xu\inst{1(}\Envelope\inst{)}}
\authorrunning{Kai Chen et al.}
\institute{School of Science, Nanjing University of Science and Technology, Nanjing, CHN\\
\and School of Computer Science and Engineering, NJUST, Nanjing, CHN\\
\email{\{kaichen,xuchung\}@njust.edu.cn}}
\maketitle
\begin{abstract}
Searchable symmetric encryption (SSE) for multi-owner model draws much attention as it enables data users to perform searches over encrypted cloud data outsourced by data owners. However, implementing secure and precise query, efficient search and flexible dynamic system maintenance at the same time in SSE remains a challenge. To address this, this paper proposes secure and efficient multi-keyword ranked search over encrypted cloud data for multi-owner model based on searching adversarial networks. We exploit searching adversarial networks to achieve optimal pseudo-keyword padding, and obtain the optimal game equilibrium for query precision and privacy protection strength. Maximum likelihood search balanced tree is generated by probabilistic learning, which achieves efficient search and brings the computational complexity close to $\mathcal{O}(\log N)$. In addition, we enable flexible dynamic system maintenance with balanced index forest that makes full use of distributed computing. Compared with previous works, our solution maintains query precision above 95\% while ensuring adequate privacy protection, and introduces low overhead on computation, communication and storage.
\end{abstract}
\section{Introduction}\label{introduction}
\paragraph{\textbf{Background and Motivation.}}
In cloud computing, searchable symmetric encryption (SSE) for multiple data owners model (multi-owner model, MOD) draws much attention as it enables multiple data users (clients) to perform searches over encrypted cloud data outsourced by multiple data owners (authorities). Unfortunately, none of the previously-known traditional SSE scheme for MOD achieve secure and precise query, efficient search and flexible dynamic system maintenance at the same time~\cite{article/Poh/2017}. This severely limits the practical value of SSE and decreases its chance of deployment in real-world cloud storage systems.
\paragraph{\textbf{Related Work and Challenge.}}
SSE has been continuously developed since it was proposed by Song et al.~\cite{proc/Song/2000}, and \textit{multi-keyword ranked search} over encrypted cloud data scheme is recognized as outstanding~\cite{article/Poh/2017}. Cao et al.~\cite{article/Cao/2014} first proposed privacy-preserving multi-keyword ranked search scheme (MRSE), and established strict privacy requirements. They first employed \textit{asymmetric scalar-product preserving encryption} (ASPE) approach~\cite{proc/Wong/2009} to obtain the similarity scores of the query vector and the index vector, so that the cloud server can return the \textit{top-k} documents. However, they did not provide the optimal balance of query precision and privacy protection strength. For better query precision and query speed, Sun et al.~\cite{article/Sun/2014} proposed MTS with the TF$\times$IDF keyword weight model, where the keyword weight depends on the frequency of the keyword in the document and the ratio of the documents containing this keyword to the total documents. This means that TF$\times$IDF cannot handle the differences between data from different owners in MOD, since each owner's data is different and there is no uniform standard to measure keyword weights. Based on MRSE, Li et al.~\cite{article/Li/2014} proposed a better solution (MKQE), where a new index construction algorithm and trapdoor generation algorithm are designed to realize the dynamic expansion of the keyword dictionary and improve the system performance. However, their scheme only realized the linean search efficiency. Xia et al.~\cite{article/Xia/2016} provided EDMRS to support flexible dynamic operation by using balanced index tree that builded following the bottom-up strategy and ``greedy" method, and they used parallel computing to improve search efficiency. However, when migrating to MOD, ordinary balanced binary tree they employed is not optimistic~\cite{article/Guo/2018}. It is frustrating that the above solutions only support SSE for single data owner model. Due to the diverse demand of the application scenario, such as emerging authorised searchable technology for multi-client (authority) encrypted medical databases that focuses on privacy protection~\cite{article/Xu/2019/1,article/Xu/2019/2}, research on SSE for MOD is increasingly active. Guo et al.~\cite{article/Guo/2018} proposed MKRS\_MO for MOD, they designed a heuristic weight generation algorithm based on the relationships among keywords, documents and owners (KDO). They considered the correlation among documents and the impact of documents' quality on search results, so that the KDO is more suitable for MOD than the TF$\times$IDF. However, they ignored the secure search scheme in \textit{known background model}~\cite{article/Cao/2014}(a \textit{threat model} that measures the ability of ``honest but curious" cloud server~\cite{proc/Wang/2010,proc/Yu/2010} to evaluate private data and the risk of revealing private information in SSE system). Currently, SSE for MOD is still these challenges: \textbf{(1)} comprehensively optimizing query precision and privacy protection is difficult; \textbf{(2)} a large amount of different data from multiple data owners make the data features sparse, and the calculation of high-dimensional vectors can cause ``curse of dimensionality"; \textbf{(3)} frequent updates of data challenge the scalability of dynamic system maintenance.
\paragraph{\textbf{Our Contribution.}}
This paper proposes secure and efficient multi-keyword ranked search over encrypted cloud data for multi-owner model based on searching adversarial networks (MRSM\_SAN). Specifically, including the following three techniques: \textbf{(1) optimal pseudo-keyword padding based on searching adversarial networks (SAN):} To improve the privacy protection strength of SSE is a top priority. Padding random noise into the data~\cite{article/Cao/2014,article/Li/2014,proc/Xu/2019} is a current popular method designed to interfere with the analysis and evaluation from cloud server, which protects the document content and keyword information better. However, such an operation will reduce the query precision~\cite{article/Cao/2014}. In response to this, we creatively use \textit{adversarial learning}~\cite{article/Goodfellow/2014} to obtain the \textit{optimal probability distribution} for controlling pseudo-keyword padding and the \textit{optimal game equilibrium} for the query precision and the privacy protection strength. This makes query precision exceeds 95\% while ensuring adequate privacy protection, which is better than traditional SSE~\cite{article/Cao/2014,article/Guo/2018,article/Li/2014,article/Sun/2014,article/Xia/2016}; \textbf{(2) efficient search based on maximum likelihood search balanced tree (MLSB-Tree):} The construction of the index tree is the biggest factor affecting the search efficiency. If the leaf nodes of the index tree are sorted by maximum probability (the ranking of the index vectors from high to low depends on the probability of being searched), the computational complexity will be close to $\mathcal{O}(\log N)$~\cite{book/Knuth/1998}. \textit{Probabilistic learning} is employed to obtain MLSB-Tree, which is ordered in a maximum probability. The experimental evaluation shows that MLSB-Tree-based search is faster and more stable compare with related works~\cite{article/Guo/2018,article/Xia/2016};
\textbf{(3) flexible dynamic system maintenance based on balanced index forest (BIF):} Using \textit{unsupervised learning}~\cite{article/Chen/2019,article/Salem/2018} to design a fast index clustering algorithm to classify all indexes into multiple index partitions, and a corresponding balanced index tree is constructed for each index partition, thus all index trees form BIF. Owing to BIF is distributed, it only needs to maintain the corresponding index partition without touching all indexes in dynamic system maintenance, which improves the efficiency of index update operations and introduces low overhead on the computation, communication and storage. In summary, MRSM\_SAN increases the possibility of deploying dynamic SSE in real-world cloud storage systems.
\paragraph{\textbf{Organization and Version Notes.}}
Section \ref{Section2} describes scheme. Section \ref{Section3} conducts experimental evaluation. Section \ref{Section4} discusses our solution. Compared with the preliminary version~\cite{article/Chen/2019/1}, this paper adds algorithms, enhances security analysis, and conducts more in-depth experimental analysis of the proposed scheme.
%The extended version~\cite{article/Chen/2019/2} provides detailed instructions and appendices.
%
\section{Secure and Efficient MRSM\_SAN}\label{Section2}
\subsection{System Model}
\begin{figure}[ht]
\small
\centering
\includegraphics[width=10cm]{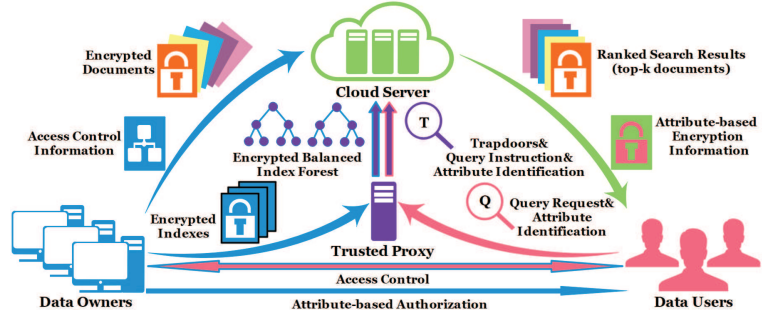}\\
\caption{The basic architecture of MRSM\_SAN}\label{fig:1}
\end{figure}

The proposed system model consists of four parties, is depicted in Fig.~\ref{fig:1}. \textbf{Data owners} ($DO$) are responsible for constructing searchable index, encrypting data and sending them to cloud server or trusted proxy; \textbf{Data users} ($DU$) are consumers of cloud services. Based on \textit{attribute-based encryption}~\cite{proc/Goyal/2006}, once $DO$ authorize $DU$ attributes related to the retrieved data, $DU$ can retrieve the corresponding data; \textbf{Trusted proxy} ($TP$) is responsible for index processing, query and trapdoor generation, user authority authentication; \textbf{Cloud server} ($CS$) provides cloud service, including running authorized access controls, performing searches over encrypted cloud data based on query requests, and returning \textit{top-k} documents to $DU$. $CS$ is considered``honest but curious"~\cite{proc/Wang/2010,proc/Yu/2010}, so that it is necessary to provide a secure search scheme to protect privacy. Our goal is to protect index privacy, query privacy and keyword privacy in dynamic SSE.
\subsection{MRSM\_SAN Framework}
\begin{description}
  \item[\textit{Setup}:] Based on index clustering results ($s$ index partitions) and privacy requirements in \textit{known background model}~\cite{article/Cao/2014}, $TP$ determines the size $N_{i}$ of sub-dictionary $D_{i}$, the number $U_{i}$ of pseudo-keyword, sets the parameter $V_{i} = U_{i} + N_{i}$. Thus $V$ = \{$V_{1}$,\ldots,$V_{s}$\}, $U$ = \{$U_{1}$,\ldots,$U_{s}$\}, $N$ = \{$N_{1}$,\ldots,$N_{s}$\}.

  \item[\textit{KeyGen}($V$):] $TP$ generates key $SK$ = \{$SK_{1}$,\ldots,$SK_{s}$\},
  where $SK_{i}$ = \{$S_{i}$, $M_{i,1}$, $M_{i,2}$\}, $M_{i,1}$ and $M_{i,2}$ are two $V_{i} \times V_{i}$-dimensional invertible matrices, $S_{i}$ is a random $V_{i}$-dimensional vector. Symmetric key $\overline{SK}_{i}$ = \{$S_{i}$, $M^{-1}_{i,1}$, $M^{-1}_{i,2}$\}.

  \item[\textit{Extended-KeyGen}($SK_{i},Z_{i}$):] For dynamic search~\cite{article/Li/2014,article/Xia/2016}, if $Z_{i}$ new keywords are added into the $i$-th sub-dictionary, $TP$ generates a new key $SK_{i}'$ = \{$S_{i}'$, $M_{i,1}'$, $M_{i,2}'$\}, where $M'_{i,1}$ and $M'_{i,2}$ are two $(V_{i}+Z_{i}) \times (V_{i}+Z_{i})$-dimensional invertible matrices, $S_{i}$ is a new random $(V_{i}+Z_{i})$-dimensional vector.

  \item[\textit{BuildIndex}($F,SK$):] To realize secure search in \textit{known background model}~\cite{article/Cao/2014}, $TP$ pads $U_{i}$ pseudo-keyword into weighted index $\overline{I}$  (associated with document $F$) to obtain secure index $\widetilde{I}$, and uses $\widetilde{I}$ and $SK$ to generate BIF $\mathcal{F}$ = \{$\tau_{1}$,\ldots,$\tau_{s}$\} and encrypted BIF $\widetilde{\mathcal{F}}$ = \{$\widetilde{\tau_{1}}$,\ldots,$\widetilde{\tau_{s}}$\}. Finally, $TP$ sends $\widetilde{\mathcal{F}}$ to $CS$.

  \item[\textit{Trapdoor}($Q,SK$):] $DU$ send query requests (keywords and their weights) and attribute identification to $TP$. $TP$ generates query $Q$ = \{$Q_{1}$,\ldots,$Q_{s}$\} and generates trapdoor $T$ = \{$T_{1}$,\ldots,$T_{s}$\} using $SK$. Finally, $TP$ sends $T$ to $CS$.

  \item[\textit{Query}$(T,\widetilde{\mathcal{F}},t,k)$:] $TP$ sends query information to $CS$ and specifies $t$ index partitions to be queried. $CS$ performs searches and retrieves \textit{top-k} documents.
  \end{description}
\subsection{Algorithms for Scheme}
%################################################################################
\begin{algorithm}[!ht]
\raggedright
\scriptsize
\caption{\textit{Binary Index Generation}}\label{alg1}
\hspace*{0in} {\bf Input:}
Document set $F = \{F_{1},\ldots,F_{m}\}$, keyword dictionary $D = \{w_{1}, \ldots, w_{n}\}.$\\
\hspace*{0in} {\bf Output:}
Binary index set $I = \{I_{1},\ldots,I_{m}\}.$\\
\begin{algorithmic}[1]
\For{$i = 1,$ \textbf{to} $m$}
\State Based on \textit{\textbf{Vector Space Model}}~\protect\cite{article/Salton/1975} and keyword dictionary $D$, $DO_{i}$ generates binary index $I_{i}$ = \{$I_{i,1}$,\ldots,$I_{i,n_{i}}$\} for document $F_{i}$ = \{$F_{i,1}$,\ldots,$F_{i,n_{i}}$\}, where $I_{i,j}$ is a binary index vector.
\State \Return Binary index Index $I$
\EndFor
\end{algorithmic}
\end{algorithm}
%################################################################################
\begin{algorithm}[!ht]
\raggedright
\scriptsize
\caption{\textit{Fast Index Clustering} {\&} \textit{Keyword Dictionary Segmentation}}\label{alg2}
\hspace*{0in} {\bf Input:}
Binary index (vector) $I$ from $DO$, where $I$ = \{$I_{1}$,\ldots,$I_{m}$\}, $DO$ = \{$DO_{1}$,\ldots,$DO_{m}$\}.\\
\hspace*{0in} {\bf Output:}
$s$ index partitions, $s$ sub-dictionaries and new binary index $\widehat{I}$ = $\{\widehat{I}_{1}, \ldots, \widehat{I}_{s}\}$.\\
\begin{algorithmic}[1]
\State \textit{Local Clustering}: For $I$ = \{$I_{1}$,\ldots,$I_{m}$\}, $TP$ uses \textit{\textbf{Twin Support Vector Machine}}~\protect\cite{article/Chen/2019} to classify the index vectors $\{I_{i,1},\ldots,I_{i,n_{i}}\}$ in $I_{i}$ into 2 clusters ($i$-th and $(m+i)$-th initial index partition) and obtain the representative vectors for the $i$-th and the $(m+i)$-th initial index partition.
\textbf{return} $2m$ initial index partitions and their representative vectors.
\State \textit{Global Clustering}: $TP$ uses \textit{\textbf{Manhattan Frequency k-Means}}~\cite{article/Salem/2018} algorithm to group all initial index partitions (representative vectors) into $s$ final index partitions.
\textbf{return} $s$ index partitions.
\State \textit{Keyword Dictionary Segmentation}: According to the obtained $s$ index partitions, the keyword dictionary $D$ is divided into $s$ sub-dictionaries $D_{1},\ldots,D_{s}$ correspondingly, where $D_{i}$ = $\{w^{i}_{1}, \ldots, w^{i}_{N_{i}}\}$. $TP$ obtains new binary index $\widehat{I} = \{\widehat{I}_{1},\ldots,\widehat{I}_{s}\}$, where $I_{i} = \{\widehat{I}_{i,1},\ldots,\widehat{I}_{i,M_{i}}\}$, $\widehat{I}_{i,j}$ is a $N_{i}$-dimensional vector. Delete ``\textbf{public redundancy zero element}" of all index vectors in the same index partition, thus the length of the index vector becomes shorter than before.
\textbf{return} $s$ sub-dictionaries and new binary index $\widehat{I}$ for $s$ index partitions.
\end{algorithmic}
\end{algorithm}
%################################################################################
\begin{algorithm}[!ht]
\raggedright
\scriptsize
\caption{\textit{Secure Weighted Index Generation}}\label{alg3}
\hspace*{0in} {\bf Input:}
Binary index $\widehat{I}$ for $s$ index partitions.\\
\hspace*{0in} {\bf Output:}
Secure weighted index $\widetilde{I}$ for $s$ index partitions, i.e. the data type is floating point.\\
\begin{algorithmic}[1]
\State \textit{Correlativity Matrix Generation}: Using the corpus to determine the semantic relationship between different keywords and obtain the \textit{correlativity matrix} $S_{N_{i}\times N_{i}}$ (symmetric matrix).
\State \textit{Weight Generation}: Based on KDO~\protect\cite{article/Guo/2018}, construct the average keyword popularity $AKP$ about $DO$. Specifically, calculate $AKP_{i}$  of $DO_{i}$ with equation ``$AKP_{i} = (P_{i}\cdot \widehat{I_{i}}) \otimes \alpha_{i}$", where the operator $\otimes$ denotes the product of two vectors¡¯ corresponding elements, $\alpha_{i}$ = ($\alpha_{i,1}$,\ldots,$\alpha_{i,N_{i}}$), if $|L_{i}(w^{i}_{t})| \neq 0$, $\alpha_{i,t}$ = $\frac{1}{|L_{i}(w^{i}_{t})|}$, otherwise $\alpha_{i,t}$ = 0 (where $|L_{i}(w^{i}_{t})|$ is the number of documents contain keyword $w^{i}_{t}$ that in $i$-th sub-dictionary $D_{i}$, $t \in \{1, \ldots, N_{i}\}$). Calculate the raw weight information for $DO_{i}$, $W^{raw}_{i}$ = $S_{N_{i}\times N_{i}} \cdot AKP_{i}$, where $W^{raw}_{i}$ = ($W^{raw}_{i,1}$,\ldots,$W^{raw}_{i,N_{i}}$).
\State \textit{Normalized Processing}: Obtain the maximum raw weight of every keyword among different $DO$, $W_{max}$ = $(W^{raw}_{i',1},W^{raw}_{i',2},\ldots)$. Based on the $W_{max}$, calculate $W_{i,t}$ = $\frac{W^{raw}_{i,t}}{W_{max}[j]}$
\State \textit{Weighted Index Generation}: $TP$ obtains weighted index vector with ``$\overline{I}_{i,j} = \widehat{I}_{i,j} \otimes W_{i}$", where $\overline{I}_{i,j}$ associated with document $F_{i,j}$ corresponds to the $i$-th index partition ($j \in\{1,\ldots,M_{i}\}$).
\State \textit{Secure Weighted Index Generation}: $TP$ pads $U_{i}$ pseudo-keyword into $\overline{I}$ in $i$-th index partition to obtain $V_{i}$-dimensional secure weighted index $\widetilde{I}$ with high privacy protection strength~\protect\cite{article/Xia/2016,proc/Xu/2019}.
\end{algorithmic}
\end{algorithm}
%################################################################################
\begin{algorithm}[!ht]
\raggedright
\scriptsize
\caption{\textit{MLSB-Tree and BIF Generation}}\label{alg4}
\hspace*{0in} {\bf Input:}
Secure weighted index $\widetilde{I}$ for $s$ index partitions, randomly generated query vector $Q$.\\
\hspace*{0in} {\bf Output:}
MLSB-Tree $\tau_{1},\ldots, \tau_{s}$ for $s$ index partitions and BIF $\mathcal{F}$ for all indexes belong to $DO$.\\
\begin{algorithmic}[1]
\For{$i = 1,$ \textbf{to} $s$}
\For{$j = 1,$ \textbf{to} $M_{i}$}
\State Based on \textbf{probabilistic learning}, $TP$ calculates ``$Score_{i,j}$ = $\sum^n_{k = 1} I^T_{i,j}\cdot Q_{i,k}$"; Then, $TP$ sorts $I_{i,1},\ldots,I_{i,M_{i}}$ according to $Score_{i,1}, \ldots, Score_{i,M_{i}}$; Finally, $TP$ follows the bottom-up strategy and generates MLSB-Tree $\tau_{i}$ (balanced tree) with \textbf{greedy} method~\cite{article/Xia/2016}.
\EndFor
\State \Return MLSB-Tree $\tau_{1},\ldots, \tau_{s}$.
\EndFor
\State \Return BIF $\mathcal{F}$ = \{$\tau_{1}$,\ldots,$\tau_{s}$\}.
\end{algorithmic}
\end{algorithm}
%################################################################################
\begin{algorithm}[!ht]
\raggedright
\scriptsize
\caption{\textit{Encrypted MLSB-Tree and Encrypted BIF Generation}}\label{alg5}
\hspace*{0in} {\bf Input:}
BIF $\mathcal{F}$ = \{$\tau_{1},\ldots, \tau_{s}$\} and key $SK = \{SK_{1},\ldots,SK_{s}\}$, where $SK_{i}$ = $\{S_{i},M_{i,1},M_{i,2}\}$.\\
\hspace*{0in} {\bf Output:}
Encrypted BIF $\widetilde{\mathcal{F}}$ = \{$\widetilde{\tau_{1}}$,\ldots,$\widetilde{\tau_{s}}$\}.\\
\begin{algorithmic}[1]
\For{$i = 1,$ \textbf{to} $s$}
\State $TP$ encrypts MLSB-Tree $\tau_{i}$ with the secret key $SK_{i}$ to obtain encrypted MLSB-Tree $\widetilde{\tau_{i}}.$
\For{$j = 1,$ \textbf{to} $n$}
\State $TP$ ``splits" vector $u_{i,j}.v$ of $u_{i,j}$ (node of $\tau_{i}$) into two random vectors $u_{i,j}.v_{1}$, $u_{i,j}.v_{2}.$
\If{$S_{i}[t]=0$}
\State $u_{i,j}.v_{1}[t]$ = $u_{i,j}.v_{2}[t]$ = $u_{i,j}.v[t].$
\Else{}
\If{$S_{i}[t]=1$}
\State $u_{i}.v_{1}[t]$ is a random value $\in (0,1)$, $u_{i,j}.v_{2}[t] = u_{i,j}.v[t]-u_{i,j}.v_{1}[t].$
\EndIf
\EndIf
\State $TP$ encrypts $u_{i,j}.v$ with reversible matrices $M_{i,1}$ and $M_{i,2}$ to obtain ``$\widetilde{u_{i,j}.v}$ = \{$\widetilde{u_{i,j}.v_{1}}$, $\widetilde{u_{i,j}.v_{2}}$\} = $\{M_{i,1}^{T}u_{i,j}.v_{1}, M_{i,2}^{T}u_{i,j}.v_{2}\}$", where $u_{i,j}.v_{1}$ and $u_{i,j}.v_{2}$ are $V_{i}$-length vectors
\EndFor
\State \Return Encrypted MLSB-Tree $\widetilde{\tau_{i}}$.
\EndFor
\State \Return Encrypted Encrypted BIF $\widetilde{\mathcal{F}}$ = \{$\widetilde{\tau_{1}}$,\ldots,$\widetilde{\tau_{s}}$\}.
\end{algorithmic}
\end{algorithm}
%################################################################################
\begin{algorithm}[!ht]
\scriptsize
\begin{multicols}{2}
\underline{\textbf{\textit{Trapdoor Generation}}}\\
\raggedright
\scriptsize
\hspace*{0in} {\bf Input:}
Query vector $Q = \{Q_{1}, \ldots, Q_{s}\}$.\\
\hspace*{0in} {\bf Output:}
Trapdoor $T = \{T_{1}, \ldots, T_{s}\}$.\\
\begin{algorithmic}[1]
\For{$i = 1,$ \textbf{to} $s$}
\State $TP$ ``splits" query vector $Q_{i}$ into two random vectors $Q_{i,1}$ and $Q_{i,2}$.
\If{$S_{i}[t]=0$}
\State $Q_{i,1}[t]$ is a random value $\in (0,1)$, $Q_{i,2}[t] = Q_{i}[t] - Q_{i,1}[t]$.
\Else{}
\If{$S_{i}[t]=1$}
\State $Q_{i,1}[t]=Q_{i,2}[t]=Q_{i}[t]$, where $t\in\{1,2,\ldots,V_{i}\}$.
\EndIf
\EndIf
\State $TP$ encrypts $Q_{i,1}$ and $Q_{i,2}$ with reversible matrices $M^{-1}_{i,1}$ and $M^{-1}_{i,2}$ to obtain trapdoor ``$T_{i}$ = $\{\widetilde{Q}_{i,1},\widetilde{Q}_{i,2}\}$ = $\{M_{i,1}^{-1}Q_{i,1}, M_{i,2}^{-1}Q_{i,2}\}$".
\EndFor
\State \Return $T = \{T_{1}, \ldots, T_{s}\}.$
\end{algorithmic}
\underline{\textbf{\textit{GDFS$(T,\widetilde{\mathcal{F}},t,k)$}}}\\
\raggedright
\scriptsize
\hspace*{0.02in} {\bf Input:}
Query$(T,\widetilde{\mathcal{F}},t,k)$.\\
\hspace*{0.02in} {\bf Output:}
\textit{top-k} documents.\\
\begin{algorithmic}[1]
\For{$i = 1,$ \textbf{to} $s$}
\If{$\tau_{i}$ is the specified index tree}
\If{$u_{i,j}$ is a non-leaf node}
\If{$Score(\widetilde{u_{i,j}.v},T_{i}) > \lceil \frac{k}{t}\rceil$-th score}
\State GDFS($u_{i,j}$.high-child)
\State GDFS($u_{i,j}$.low-child)
\Else{}
\State \Return
\EndIf{}
\Else{}
\If{$Score(\widetilde{u_{i,j}.v},T_{i}) > \lceil\frac{k}{t}\rceil$-th score}
\State Update $\lceil\frac{k}{t}\rceil$-th score for $i$-th index tree $\tau_{i}$ and the ranked search result list for $\widetilde{\mathcal{F}}$.
\EndIf
\EndIf
\EndIf
\EndFor
\State \Return The final \textit{top-k} documents for $\widetilde{\mathcal{F}}.$
\end{algorithmic}
\end{multicols}
\caption{\textit{Trapdoor Generation and GDFS$(T,\widetilde{\mathcal{F}},t,k)$}}\label{alg6}
\end{algorithm}
%################################################################################
\begin{enumerate}
  \item \textbf{\textit{Binary Index Generation}:} $DO_{i}$ uses algorithm~\ref{alg1} to generate the binary index (vector) $I_{i}$ for the document $F_{i}$, and sends $I_{i}$ to $TP$.

  \item \textbf{\textit{Fast Index Clustering} {\&} \textit{Keyword Dictionary Segmentation}:} We employ algorithm~\ref{alg2} to solve ``curse of dimensionality" issue in computing.

  \item \textbf{\textit{Weighted Index Generation}:} $TP$ exploits the KDO weight model~\cite{article/Guo/2018} to generate the weighted index, as shown in algorithm~\ref{alg3}.

  \item \textbf{\textit{MLSB-Tree and BIF Generation}:} $TP$ uses algorithm~\ref{alg4} to generate MLSB-Tree $\tau_{1}, \ldots, \tau_{s}$ and BIF $\mathcal{F}$ = $\{\tau_{1}, \ldots, \tau_{s}\}$.
%\end{enumerate}
%\begin{enumerate}
  \item \textbf{\textit{Encrypted MLSB-Tree and Encrypted BIF Generation}.} $TP$ encrypts $\mathcal{F}$ using algorithm~\ref{alg5} and sends encrypted $\widetilde{\mathcal{F}}$ to $CS$. $\tau_{i}$ and $\widetilde{\tau_{i}}$ are isomorphic (i.e.$\tau_{i} \cong \widetilde{\tau_{i}}$)~\cite{article/Xia/2016}. Thus, the search capability of tree is still well maintained.

  \item\textbf{\textit{Trapdoor Generation}.} Based on query request from $DU$, $TP$ generates $Q$ = \{$Q_{1}$,\ldots,$Q_{s}$\} and $T$ = \{$T_{1}$,\ldots,$T_{s}$\} using algorithm~\ref{alg6}, and sends $T$ to $CS$.

\item\textbf{\textit{Search Process}.} \textbf{(1)} \textit{Query Preparation}: $DU$ send query request and attribute identifications to $TP$. If validating queries are valid, $TP$ generates trapdoors and initiates search queries to $CS$. If access control passes, $CS$ performs searches and returns \textit{top-k} documents to $DU$. Otherwise $CS$ refuses to query. \textbf{(2)} \textit{Calculate Matching Score for Query on MLBS-Tree $\tau_{i}$}:
\begin{eqnarray*}
\scriptstyle
Score(\widetilde{u_{i,j}.v} ,T_{i}) = \{M_{i,1}^{T}u_{i,j}.v_{1}, M_{i,2}^{T}u_{i,j}.v_{2}\} \cdot \{M_{i,1}^{-1}Q_{i,1}, M_{i,2}^{-1}Q_{i,2}\} = u^{T}_{i,j}.v \cdot Q_{i}
\end{eqnarray*}
\textbf{(3)} \textit{Search Algorithm for BIF}: the greedy depth-first search (GDFS) algorithm for BIF as shown in algorithm~\ref{alg6}.
\end{enumerate}
\subsection{Security Improvement and Analysis}
\begin{figure}[ht]
\small
\centering
\includegraphics[width=8cm]{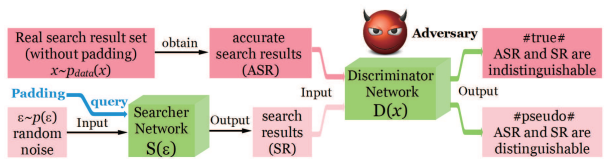}
\caption{Searching Adversarial Networks}\label{fig:2}
\end{figure}

\paragraph{\textbf{Adversarial Learning.}}
Padding random noise into the data~\cite{article/Cao/2014,article/Li/2014,proc/Xu/2019} is a popular method to improve security. However, pseudo-keyword padding that follows different probability distributions will reduce query precision to varying degrees~\cite{article/Cao/2014,article/Li/2014}. Therefore, it is necessary to optimize the probability distribution that controls pseudo-keyword padding. To address this, \textit{adversarial learning}~\cite{article/Goodfellow/2014} for optimal pseudo-keyword padding is proposed. As shown in Fig.~\ref{fig:2}. \textbf{\textit{Searcher Network S($\varepsilon$)} :} The search result (SR) is generated by taking the random noise $\varepsilon$ ($\varepsilon \sim p(\varepsilon)$, $p(\varepsilon)$ is the object probability distribution ) as an input and performing a search, and supplies SR to the discriminator network $D(x)$. \textbf{\textit{Discriminator Network D($x$)}:} The input has an accurate search result (ASR) or SR and attempts to predict whether the current input is an ASR or a SR. One of the inputs $x$ is obtained from the real search result set distribution $p_{data}(x)$, and then one or two are solved. Classify problems and generate scalars ranging from 0 to 1. Finally, in order to reach a balance point which is the best point of the minimax game, $S(\varepsilon)$ generates SR, and $D(x)$ (considered as adversary) considers the probability that $S(\varepsilon)$ produces ASR is $0.5$, i.e. it is difficult to distinguish between padding and without-padding, thus it can achieve effective security~\cite{proc/Xu/2019}.

\noindent Similar to GAN~\cite{article/Goodfellow/2014}, to learn the searcher's distribution $p_{s}$ over data $x$, we define a prior on input noise variables $p_{\varepsilon}(\varepsilon)$, then represent a mapping to data space as $S(\varepsilon; \theta_{s})$, where $S$ is a differentiable function represented by a multi-layer perception with parameters $\theta_{g}$. We also define a second multi-layer perception $D(x; \theta_{d})$ that outputs a single scalar. $D(x)$ represents the probability that $x$ came from the data rather than $p_{s}$. We train $D$ to maximize the probability of assigning the correct label to both training examples and samples from $S$. We simultaneously train $S$ to minimize $\log(1 - D(S(\varepsilon)))$: In other words, $D$ and $S$ play the following two-player minimax game with value function $V(S, D)$:
\begin{equation*}
\min_{S}\max_{D}V(D,S)= \mathbb{E}_{x\sim p_{data}(x)}[\log D(x)] + \mathbb{E}_{x\sim p_{\varepsilon}(\varepsilon)}[\log (1 - D(S(\varepsilon)))]
\end{equation*}
\paragraph{\textbf{Security Analysis.}}
\textit{Index confidentiality and query confidentiality:} ASPE approach~\cite {proc/Wong/2009} is widely used to generate secure index/query in privacy-preserving keyword search schemes~\cite{article/Cao/2014,article/Guo/2018,article/Li/2014,article/Sun/2014,article/Xia/2016} and its security has been proven. Since the index/query vector is randomly generated and search queries return only the secure inner product~\cite{article/Cao/2014} computation results (non-zero) of encrypted index and trapdoor, thus $CS$ is difficult to accurately evaluate the keywords including in the query and matching \textit{top-k} documents. Moreover, confidentiality is further enhanced as the optimal pseudo-keyword padding is difficult to distinguish and the transformation matrices are harder to figure out~\cite{proc/Wong/2009}.

\textit{Query unlinkability:} By introducing the random value $\varepsilon$ (padding pseudo-keyword), the same search requests will generate different query vectors and receive different relevance score distributions~\cite{article/Cao/2014,article/Xia/2016}. The optimal game equilibrium for precision and privacy is obtain by \textit{adversarial learning}, which further improves query unlinkability. Meanwhile, SAN are designed to protect access pattern~\cite{proc/Xu/2019}, which makes it difficult for $CS$ to judge whether the retrieved ranked search results come from the same request.

\textit{Keyword privacy:}  According to the security analysis in~\cite{article/Xia/2016}, for $i$-th index partition, aiming to maximize the randomness of the relevance score distribution, it is necessary to obtain as many different $\sum \varepsilon_{\nu_{i}}$ as possible (where $\nu_{i} \in \{j|Q_{i}[j+N_{i}]=\alpha_{i}, j = 1,\ldots,U_{i}\}$; in~\cite{article/Xia/2016}, $\alpha_{i} = 1$). Assuming each index vector has at least $2^{\omega_{i}}$ different $\sum \varepsilon_{\nu_{i}}$ choices, the probability of two $\sum \varepsilon_{\nu_{i}}$ share the same value is less than $\frac{1}{2^{\omega_{i}}}$. If we set each $\varepsilon_{j} \sim U(\mu_{i}'-\delta_{i}, \mu_{i}'+\delta_{i})$ (\textit{Uniform distribution}), according to the central limit theorem, $\sum \varepsilon_{\nu_{i}} \sim N(\mu_{i}, \sigma^2_{i})$ (\textit{Normal distribution}), where $\mu_{i} = \omega_{i} \mu_{i}'$, $\sigma^2 = \frac{\omega_{i} \delta^2_{i}}{3}$. Therefore, it can set $\mu_{i} = 0$ and balance precision and privacy by adjusting the variance $\sigma^{2}_{i}$ in real-world application. In fact, when $\alpha_{i} \in [0,1]$ (floating point number), SAN can achieve stronger privacy protection.

\section{Experimental Evaluation}\label{Section3}
\begin{figure}[ht]
\centering
\makeatletter\def\@captype{figure}\makeatother
\subfigure[]{\includegraphics[width=0.38\textwidth]{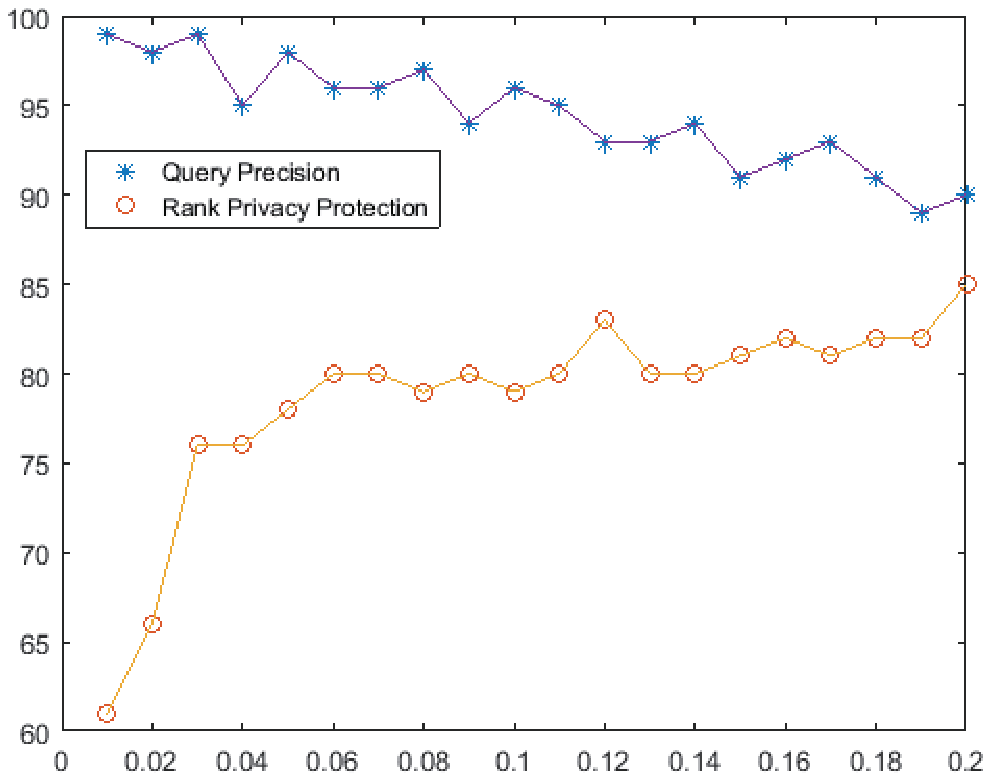}\label{fig:3a}}
\hfil
\subfigure[]{\includegraphics[width=0.375\textwidth]{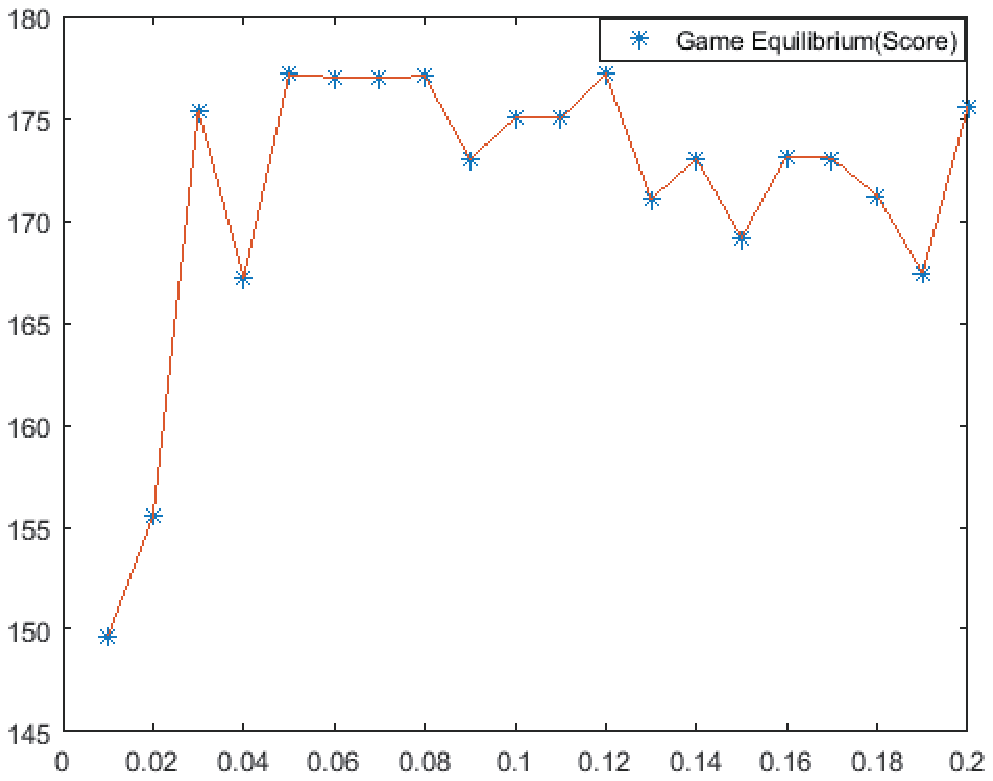}\label{fig:3b}}
\caption{With different choice of standard deviation $\sigma$ for the random variable $\varepsilon$. \scriptsize \{\textbf{(a)} query precision(\%) and rank privacy protection(\%); \textbf{(b)} game equilibrium (score). explanation for $\sigma \in [0.01, 0.2]$: When $\sigma$ is greater than 0.2, the weight of the pseudo-keyword may be greater than 1, which violates our weight setting (between 0 and 1),  so we only need to find the best game equilibrium point when $\sigma \in [0.01, 0.2]$.)}
\label{fig:3}
\end{figure}

We implemented the proposed scheme using Python in Windows 10 operation system with Intel Core i5 Processor 2.40GHz and evaluated its performance on a real-world data set (academic conference publications provided by IEEE xplore \url{https://ieeexplore.ieee.org/}, including 20,000 papers and 80,000 different keywords, 400 academic conferences were randomly selected as data owners $DO$). All experimental results represent the average of 1000 trials.
\paragraph{\textbf{Optimal Pseudo-keyword Padding.}}
The parameters controlling the probability distribution (using SAN to find or approximate) are adjusted to find the optimal game equilibrium for \textit{query precision} $P_{k}$ (denoted as $x$) and \textit{rank privacy protection} $P_{k}'$ (denoted as $y$) (where $P_{k} = k'/k$, $P_{k}' = \sum |r_{i}-r_{i}'|/k^2$, $k'$ and $r_{i}$ are respectively the number of real \textit{top-k} documents and the rank number of document in the retrieved $k$ documents, and $r_{i}'$ is document's real rank number in the whole ranked results~\cite{article/Cao/2014}). We choose 95\% query precision and 80\% rank privacy protection as benchmarks to get the \textit{game equilibrium score} calculation formula: $f(x,y) = \frac{1}{95}x^2 + \frac{1}{80}y^2$ (objective function to be optimized). As shown in Fig.~\ref{fig:3}, we find the optimal game equilibrium ($\max f(x,y) = 177.5$) at $\sigma_{1} = 0.05$, $\sigma_{2} =0.08$, $\sigma_{3} =0.12$. The corresponding query precision are: 98\%, 97\%, 93\%. The corresponding rank privacy protection are: 78\%,79\%,84\%. Therefore, we can choose the best value of $\sigma$ to achieve optimal pseudo-keyword padding to satisfy query precision requirement and maximize rank privacy protection.
\begin{figure}[ht]
\centering
\makeatletter\def\@captype{figure}\makeatother
\subfigure[]{\includegraphics[width=0.38\textwidth]{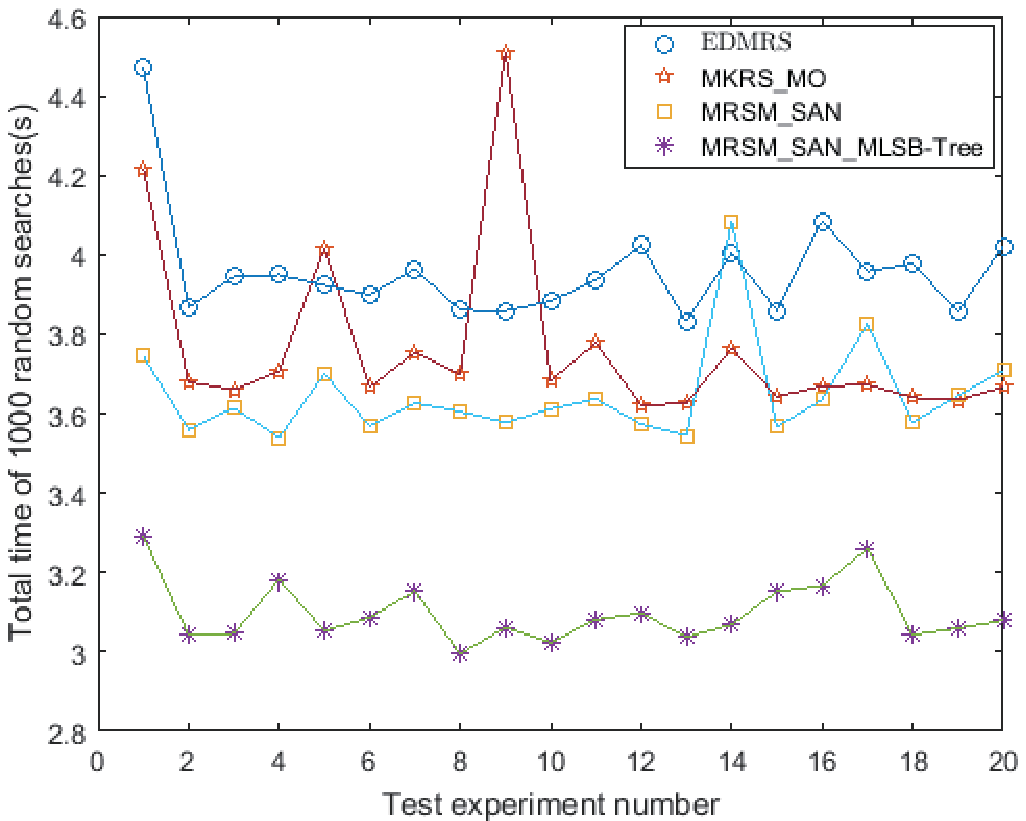}\label{fig:4a}}
\hfil
\subfigure[]{\includegraphics[width=0.375\textwidth]{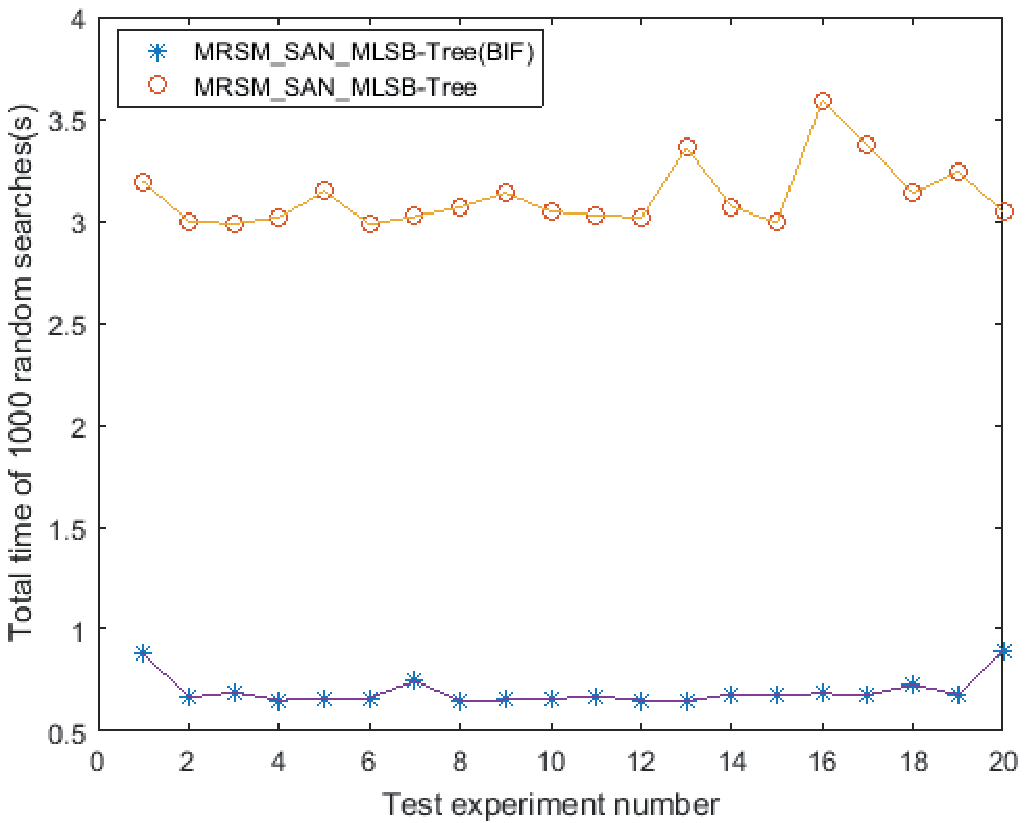}\label{fig:4b}}
\caption{Time cost of query for 1000 random searches in 500 sizes of data set. \scriptsize{\textbf{(a)} Comparison of tree-based search efficiency. Since the query is random, the search time fluctuates, which causes the curves in the graph to have intersections; \textbf{(b)} Comparison of MLSB-Tree and BIF search efficiency.}}
\label{fig:4}
\end{figure}
\paragraph{\textbf{Search Efficiency of MLSB-Tree.}}
Search efficiency is mainly described by query speed, and our experimental objects are index trees that are structured with different strategy: EDMRS~\cite{article/Xia/2016} (ordinary balanced binary tree), MKRS\_MO~\cite{article/Guo/2018} (grouped balanced binary tree), MRSM\_SAN(globally grouped balanced binary tree) and MRSM\_SAN\_MLSB-Tree. We first randomly generate 1000 query vectors, then perform search on each index tree respectively, finally take the results of 20 repeated experiments for analysis. As shown in Fig.~\ref{fig:4a}, the query speed and query stability based on MLSB-Tree are better than other index trees. Compared with EDMRS and MKRS\_MO, query speed increased by 21.72\% and 17.69\%. In terms of stability, MLSB-Tree is better than other index trees. (variance of search time(s): 0.0515~\cite{article/Guo/2018}, 0.0193~\cite{article/Xia/2016}, \textbf{0.0061}[MLSB-Tree])
\paragraph{\textbf{Search Efficiency of BIF.}}
As shown in Fig.~\ref{fig:4b}, query speed of MRSM\_SAN (with MLSB-Tree and BIF) is significantly higher than MRSM\_SAN (only with MLSB-Tree), and the search efficiency is improved by 5 times and the stability increase too. This is just the experimental result of 500 documents set with  the 4000-dimensional keyword dictionary. After the index clustering operation, the keyword dictionary is divided into four sub-dictionaries with a dimension of approximately 1000. As the amount of data increases, the dimension of the keyword dictionary will become extremely large, and the advantages of BIF will become more apparent. In our analytical experiments, the theoretical efficiency ratio before and after segmentation is: $\eta = s \frac{\mathcal{O}(\log N)}{\mathcal{O}(\log N)-\mathcal{O}(\log s)}$,where $s$ denotes the number of index partitions after fast index clustering, and $N$ denotes the total number of documents. When the amount of data increases to 20,000, the total keyword dictionary dimension is as high as 80,000. If the keyword sub-dictionary dimension is 1000, the number of index partitions after fast index clustering is 80, the search efficiency will increase by more than 100 times ($\eta = 143$). This will bring huge benefits to large information systems, and our solutions can exchange huge returns with minimal computing resources.
\begin{figure}[ht]
\centering
\makeatletter\def\@captype{figure}\makeatother
\subfigure[]{\includegraphics[width=0.4\textwidth]{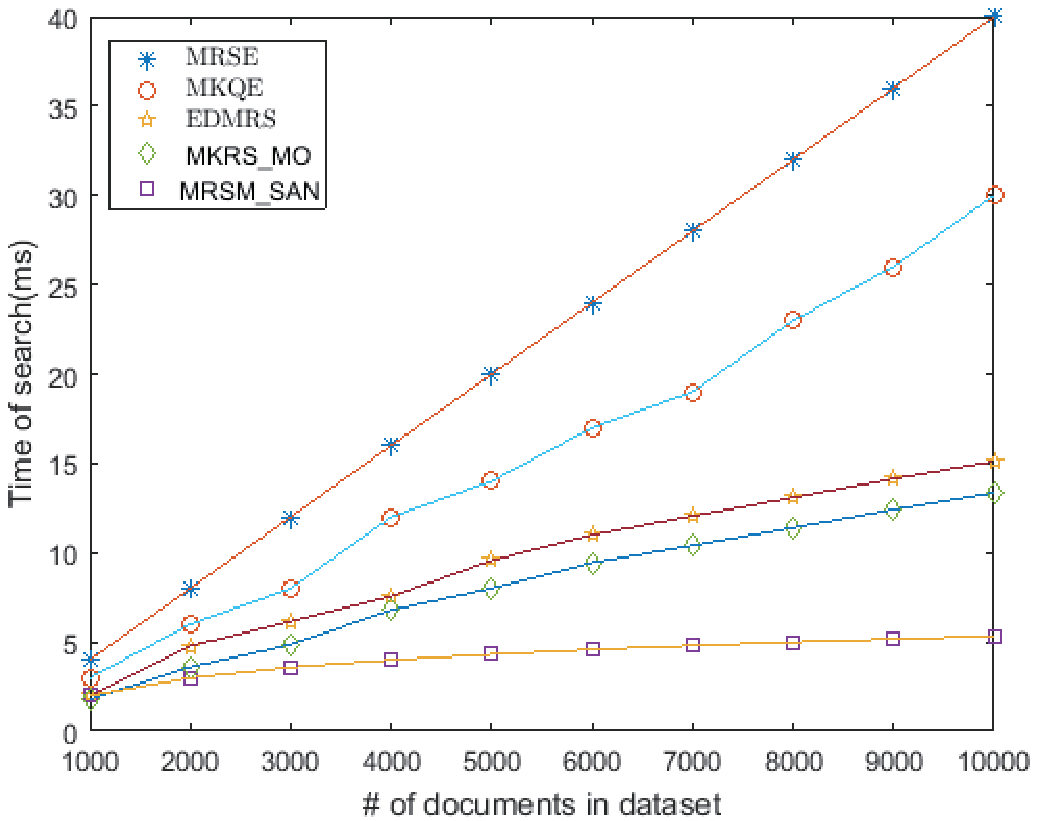}\label{fig:5a}}
\hfil
\subfigure[]{\includegraphics[width=0.38\textwidth]{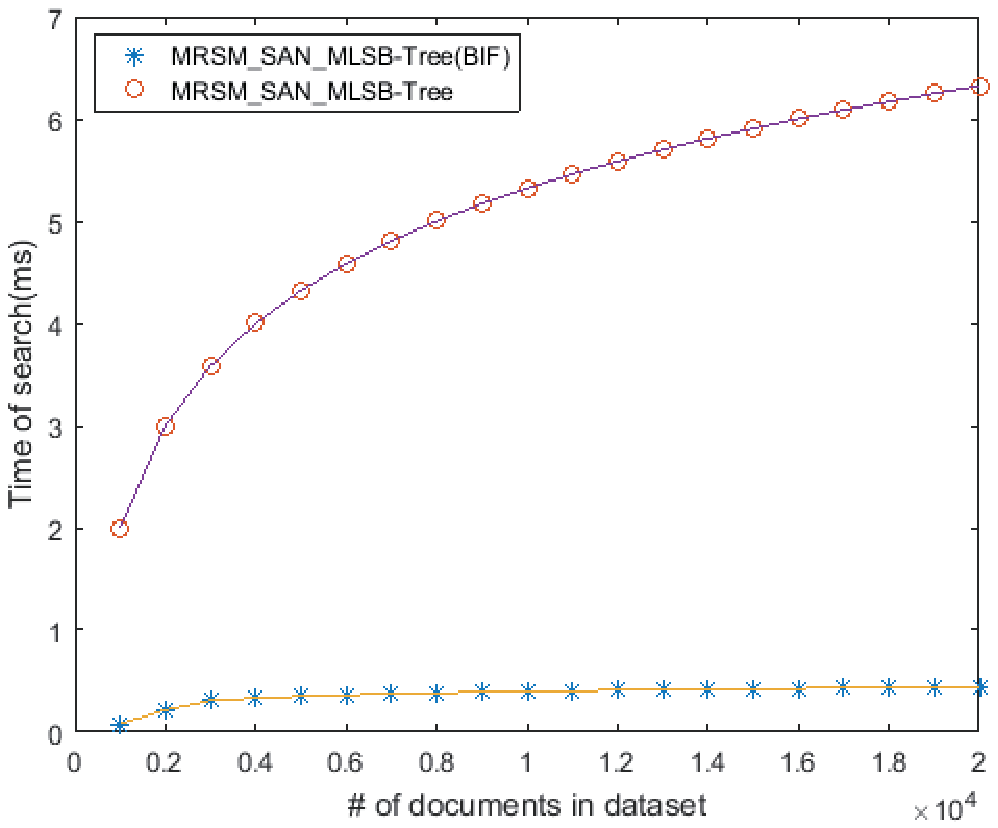}\label{fig:5b}}
\caption{Time cost of query for the same query keywords (10 keywords) in different sizes of data set.\scriptsize \{\textbf{(a)} Experimental results show that our solution achieves near binary search efficiency and is superior to other existing comparison schemes. As the amount of data increases, our solution has a greater advantage. It is worth noting that this is just the search performance based on MLSB-Tree; \textbf{(b)} Comparison of MLSB-Tree and BIF. Based on experimental analysis, it concludes that when data volume grows exponentially data features become more sparse, if all index vectors only rely on an index tree to complete the search task, the computational complexity will be getting farther away from $\mathcal{O}(\log N)$. Sparseness of data features makes the similarity between index vectors is mostly close to zero or even equal to zero, which brings trouble to the pairing of index vectors. Moreover,  the construction of balanced index tree is not global order, so it is necessary to traverse  many nodes in the search, which proves the limitation of \textit{balanced binary tree}~\protect\cite{article/Guo/2018,article/Xia/2016}. We construct MLSB-Tree with \textit{maximum likelihood method} and \textit{probabilistic learning}. Interestingly£¬the closer the number of random searches is to infinity, the higher the search efficiency of obtained index tree, this makes the computational complexity of search can converge to $\mathcal{O}(\log N)$.\}}
\label{fig:5}
\end{figure}
\paragraph{\textbf{Comparison of Search Efficiency (Larger Data Set).}}
The efficiency of MRSM\_SAN (without BIF) and related works~\cite{article/Cao/2014,article/Guo/2018,article/Li/2014,article/Xia/2016} are show as Fig.~\ref{fig:5a}, and the efficiency of MRSM\_SAN(without BIF) and MRSM\_SAN(with BIF) are show as Fig.~\ref{fig:5b}. It is more notable that the maintenance cost of scheme based on BIF is much lower than the cost of scheme only based on a balanced index tree. When adding a new document to $CS$, we need to insert a new index vector (the node of the tree) in the index tree accordingly. If it is only based on an index tree, search complexity (search the location where the new index is inserted into the index tree) and update complexity (update the parent node corresponding to the new index) are both at least $\mathcal{O}(\log N)$~\cite{article/Poh/2017}, the total cost is $2\mathcal{O}(\log N)$ (where $N$ denotes the total number of documents). But BIF is very different, because we group all index vectors into $s$ different index partitions and reduce their dimension. We assume that the number of index vectors in each index partition is equal, thus we need to spend the same update operation for each partition, which makes the cost is only $\frac{2}{s}\mathcal{O}(\log \frac{N}{s})$ and enables flexible dynamic system maintenance.  Moreover, the increase in efficiency is positively correlated with the increase in data volume and data sparsity. For communication, compared with traditional SSE schemes~\cite{article/Guo/2018,article/Sun/2014,article/Xia/2016}, when the old index tree in the cloud is overwritten by the new index tree uploaded by $DO$/$TP$, our scheme only needs to update the specified index tree instead of the entire index forest. For storage, tree-based overhead is $(2^{\log N} - \frac{1}{2})/(2^{\log \frac{N}{s}} - \frac{1}{2}) \approx 2^{\log s}$ times forest-based.
%tree-based needs $\frac{n}{2} (2^{\log N} - \frac{1}{2})$, forest-based needs $\frac{n}{2} (2^{\log \frac{N}{s}} - \frac{1}{2})$.

\section{Discussion}\label{Section4}
This paper proposes secure and efficient MRSM\_SAN, and conducts in-depth security analysis and experimental evaluation. Creatively using \textit{adversarial learning} to find \textit{optimal game equilibrium} for query precision and privacy protection strength and combining traditional SSE with \textit{uncertain control theory}, which opens a door for \textit{intelligent SSE}. In addition, we propose MLSB-Tree, which generated by a sufficient amount of random searches and brings the computational complexity close to $\mathcal{O}(\log N)$. It means that using \textit{probabilistic learning} to optimize the query result is effective in an \textit{uncertain system} (owner's data and user's queries are uncertain). Last but not least, we implement flexible dynamic system maintenance with BIF, which not only reduces the overhead of dynamic maintenance and makes full use of distributed computing, but also improves the search efficiency and achieves fine-grained search. This is beneficial to improve the availability, flexibility and efficiency of dynamic SSE system.
\section*{Acknowledgment}
This work was supported by ``the Fundamental Research Funds for the Central Universities" (No. 30918012204) and ``the National Undergraduate Training Program for Innovation and Entrepreneurship" (Item number: 201810288061). NJUST graduate Scientific Research Training of `Hundred, Thousand and Ten Thousand' Project ``\textit{Research on Intelligent Searchable Encryption Technology}".
\bibliographystyle{splncs04}
\bibliography{mybibliography}
\end{document}